\documentclass[11pt,prd,preprintnumbers,nofootinbib,superscriptaddress,notitlepage]{revtex4-1}

\usepackage{graphics}
\usepackage{bm}
\usepackage{cancel}
\usepackage{color}
\usepackage{xcolor}
\usepackage{multirow}
\usepackage{slashed}
\usepackage{array}
\usepackage{amssymb}
\usepackage{bbold}
\usepackage{graphicx}
\usepackage{mathrsfs}
\usepackage{diagbox}
\usepackage{amsmath}
\usepackage{float}
\usepackage{extarrows}
\usepackage{booktabs}
\usepackage[colorlinks,citecolor=blue,anchorcolor=red,menucolor=red,linkcolor=red,filecolor=red,runcolor=red,urlcolor=red,frenchlinks=red]{hyperref}
\usepackage{subfigure}
\usepackage{ulem}

\makeatletter
\newcommand{\thickhline}{\noalign {\ifnum 0=`}\fi \hrule height 1pt\futurelet \reserved@a \@xhline}
\newcolumntype{"}{@{\hskip\tabcolsep\vrule width 1pt\hskip\tabcolsep}}
\makeatother

\begin{document}
\title{Heavy Higgs effects on top-quark spin correlations and quantum entanglement at polarized photon colliders}

\author{Xin-Yi Zhang}
\email{sps\_zhangxy@ujn.edu.cn}
\affiliation{School of Physics and Technology, University of Jinan, Jinan, Shandong 250022, China}

\author{Hong-Xin Wang}
\email{hxwang@ytu.edu.cn}
\affiliation{Department of Physics, Yantai University, Yantai, Shandong 264005, China}

\author{Hong-Lei Li}
\email{sps\_lihl@ujn.edu.cn}
\affiliation{School of Physics and Technology, University of Jinan, Jinan, Shandong 250022, China}

\author{Tai-Fu Feng}
\email{fengtf@hbu.edu.cn}
\affiliation{Department of Physics, Hebei University, Baoding 071002, China}
\affiliation{Hebei Key Laboratory of High-precision Computation and Application of Quantum Field Theory, Baoding 071002, China}
\affiliation{Hebei Research Center of the Basic Discipline for Computational Physics, Baoding 071002, China}

\author{Zong-Guo Si}
\email{zgsi@sdu.edu.cn}
\affiliation{School of Physics, Shandong University, Jinan, Shandong 250100, China}

\begin{abstract}

We investigate the effects of a heavy CP-even Higgs boson $H$ in $\gamma\gamma\to t\bar t$ within the Type-II two-Higgs-doublet model (2HDM), including its interference with the Standard Model (SM) contribution. 
After checking the theoretical constraints and confronting the benchmark point with current experimental limits, we study the cross section, spin correlations, and quantum entanglement for different initial-state photon polarization configurations. 
While the heavy Higgs contribution has a limited impact on the inclusive cross section, it modifies the invariant-mass distribution and selected spin-correlation coefficients, leading to changes in the entanglement properties of the $t\bar t$ system, as reflected in the concurrence and Bell--CHSH parameter. 
These effects are generally more pronounced than those induced by the SM next-to-leading-order (NLO) QCD corrections.
These results demonstrate that spin correlations and quantum-entanglement observables provide complementary probes of heavy-Higgs effects beyond conventional production-rate measurements at polarized photon colliders.

\end{abstract}
\maketitle

\section{Introduction}
\label{sec:introduction}
SM provides a remarkably successful description of particle physics, yet it cannot account for several observed phenomena, motivating searches for physics beyond the Standard Model (BSM).
The top quark, as the heaviest fermion in the SM, has a large Yukawa coupling and provides an important probe of electroweak symmetry breaking and potential new physics effects~\cite{Bernreuther:2008ju,Beneke:2000hk,Chakraborty:2003iw}.
The top quark decays before hadronization, allowing its spin information to be accessed through the angular distributions of its decay products.
The $t\bar t$ system therefore provides a natural probe of two-particle spin states and quantum correlations in high-energy scattering.

The spin properties of the $t\bar t$ system have been studied through spin-polarization and spin-correlation observables~\cite{Mahlon:2010gw,Bernreuther:2001rq,Bernreuther:2013aga,Bernreuther:2015yna}.
The $t\bar t$ spin density matrix also provides a framework for characterizing quantum correlations, including quantum entanglement~\cite{Horodecki:2009zz,Gigena:2017agv,Afik:2020onf,Fabbrichesi:2021npl}.
Such correlations have recently been measured at the LHC.
ATLAS and CMS have reported measurements of top-quark spin correlations consistent with the SM predictions~\cite{ATLAS:2014abv,CMS:2019nrx}.
ATLAS measured the entanglement marker $D=-0.537\pm0.002~(\mathrm{stat})\pm0.019~(\mathrm{syst})$ for $340~\mathrm{GeV}<m_{t\bar t}<380~\mathrm{GeV}$, finding a deviation of more than $5\sigma$ from the no-entanglement hypothesis~\cite{ATLAS:2023fsd}.
CMS subsequently measured $D=-0.480^{+0.026}_{-0.029}$~\cite{CMS:2024pts}.
These measurements establish quantum correlations in top-quark pair production as experimentally accessible observables.
The $t\bar t$ spin correlations provide additional sensitivity to BSM interactions, which can modify the production amplitudes for different spin configurations and alter the spin density matrix and its associated quantum correlations~\cite{Bernreuther:2017yhg,Afik:2022dgh,Aoude:2022imd,Arai:2026jtc,Erdogan:2026cbe}.
The sensitivity of quantum-entanglement observables to BSM effects has been investigated in several recent studies~\cite{Maltoni:2024tul,Fabbrichesi:2022ovb,Subba:2026nzs,Aoude:2025jzc}.
These studies motivate the use of quantum observables as complementary probes of new interactions beyond conventional production rates and kinematic distributions.

Photon colliders provide a complementary environment for studying top-quark spin correlations and their sensitivity to BSM interactions. 
In particular, the polarization of the backscattered photons can be controlled through the initial electron and laser beam polarizations, allowing different initial-state configurations to be explored in $\gamma\gamma\to t\bar t$~\cite{Ginzburg:1981vm,Ginzburg:1982yr,HARLANDER1995137}.
The process $\gamma\gamma\rightarrow t\bar t$ has been extensively studied in terms of production rates, differential distributions, and threshold behavior~\cite{Jikia:1996bi,Denner:1995ar,Melles:1998xd,Penin:1998mx,Czarnecki:2001gi,Capatti:2025khs}.
More recently, spin correlations and quantum-entanglement observables in this process have also been investigated, including leading order (LO)~\cite{Choi:2026omc} and NLO QCD corrections~\cite{Jiang:2026sfw}.
However, the effects of a heavy Higgs boson and its interference with the SM on the $t\bar t$ spin state and quantum correlations have not been systematically studied.
The 2HDM provides a concrete framework for investigating these effects~\cite{Branco:2011iw}.
In the Type-II 2HDM, the heavy Higgs boson can contribute to $\gamma\gamma\rightarrow t\bar t$ through the $s$-channel process.
Its amplitude interferes with the SM contribution, modifying the relative contributions of different top-quark spin configurations and hence the $t\bar t$ spin density matrix.

In this work, we investigate the effects of the heavy CP-even Higgs boson in the Type-II 2HDM on \(\gamma\gamma\rightarrow t\bar t\) in polarized photon collisions, considering the interference between the Higgs-mediated amplitude and the SM.
We study their effects on the $m_{t\bar t}$ distribution and the $t\bar t$ spin density matrix, including spin correlations and quantum entanglement for different photon polarization configurations.
By comparing these quantum observables with conventional production observables, we assess the sensitivity of $t\bar t$ quantum correlations to heavy-Higgs effects.
The paper is organized as follows: Sec.~\ref{sec2} presents the theoretical framework, Sec.~\ref{sec4} presents the numerical results, and Sec.~\ref{sec5} summarizes our conclusions.


\section{Theoretical Framework}
\label{sec2}

In a photon collider setup, high-energy photon beams are generated through Compton backscattering of laser photons off energetic electron beams~\cite{  Ginzburg:1982yr}. 
Consequently, the observable cross section for top-quark pair production is obtained by convoluting the partonic cross section with the energy spectra of the backscattered photons,
\begin{equation}
\sigma_{t\bar t}=
\int_0^{y_{\rm max}}dy_1
\int_0^{y_{\rm max}}dy_2\,
f_\gamma^e(y_1,P_e^{(1)},P_L^{(1)})
f_\gamma^e(y_2,P_e^{(2)},P_L^{(2)})
\,
\hat{\sigma}_{\gamma\gamma\to t\bar t}
(\hat s,\lambda_1,\lambda_2),
\end{equation}
where $y_i$ denotes the fraction of the electron-beam energy carried by the $i$-th backscattered photon, and $\hat s=y_1y_2s_{ee}$ is the invariant mass squared of the photon--photon subsystem.

The normalized energy spectrum of the backscattered photons is determined by the Compton scattering dynamics and is given by~\cite{Ginzburg:1981vm}
\begin{equation}
f_\gamma^e(y,P_e,P_L)
=
{\cal N}^{-1}
\left[
\frac{1}{1-y}
-y
+(2r-1)^2
-
P_eP_L\,xr(2r-1)(2-y)
\right],
\end{equation}
with
\begin{equation}
r=\frac{y}{x(1-y)},
\qquad
x=\frac{4E_LE_e}{m_e^2}.
\end{equation}
Here, $E_e$ and $E_L$ denote the energies of the electron beam and laser beam, respectively, while $m_e$ is the electron mass. 
The quantities $P_e$ and $P_L$ represent the polarization of the electron beam and laser photons respectively. 
The normalization factor ${\cal N}$ satisfies
\begin{equation}
\int_0^{y_{\rm max}}
f_\gamma^e(y,P_e,P_L)\,dy=1.
\end{equation}
The kinematically allowed range of the photon energy fraction is
$0\le y\le x/(x+1)$.
To avoid $e^+e^-$ pair creation in the interaction between the backscattered and laser photons, the parameter $x$ is usually limited to $x_{\rm max}=2(1+\sqrt2)$.
For an initial $e^+e^-$ center-of-mass energy of $\sqrt{s_{ee}}=1~\mathrm{TeV}$, this gives a maximum photon--photon invariant mass of $\sqrt{\hat{s}_{\rm max}}\approx 828.4~\mathrm{GeV}$.

The Compton backscattering dynamics also determines the polarization of the scattered photons.
The  polarization of the incoming photons entering the hard-scattering subprocess is defined by
\begin{equation}
\lambda_i=
P_\gamma(y_i,P_e^{(i)},P_L^{(i)}),
\qquad i=1,2,
\end{equation}
where the function $P_\gamma(y,P_e,P_L)$ describes the polarization degree of a backscattered photon carrying an energy fraction $y$~\cite{Ginzburg:1981vm}:
\begin{equation}
P_\gamma(y,P_e,P_L)=
\frac{1}
{{\cal N}f_\gamma^e(y,P_e,P_L)}
\left\{
xrP_e
\left[
1+(1-y)(2r-1)^2
\right]
-(2r-1)P_L
\left[
\frac{1}{1-y}+1-y
\right]
\right\}.
\end{equation}

For the hard-scattering subprocess, we consider
\begin{equation}
\gamma(p_1,\lambda_1)+\gamma(p_2,\lambda_2)
\to t(k_1,s_t)+\bar{t}(k_2,s_{\bar t}),
\end{equation}
where $s_t$ and $s_{\bar{t}}$ denote the spin four-vectors of the top quark and antiquark, and they satisfy the following relation~\cite{Brandenburg:2005uu},
\begin{equation}
s_t^2=s_{\bar{t}}^2=-1, \qquad
k_1\cdot s_t = k_2\cdot s_{\bar{t}}=0.
\end{equation}

For fixed photon helicities, the polarized partonic differential cross section is written as
\begin{equation}
d\hat{\sigma}(\lambda_1,\lambda_2) =\frac{1}{2\hat{s}}|\mathcal M_{\alpha \beta}(\lambda_1,\lambda_2)|^2d\Phi_2,
\end{equation}
where  $d\Phi_2$ is two-body phase space, the indices $\alpha$ and $\beta$ label the spin states of the top quark and antiquark, respectively.
The spin dependence of the squared matrix element can be decomposed into unpolarized, single-spin polarization, and spin-correlation contributions.
In the type-II 2HDM, we consider the contribution of the heavy Higgs boson through the $s$-channel process.
Including both the SM background and the heavy Higgs contribution, the squared matrix element $|{\cal M_{\alpha \beta}}|^2$ can be written as
\begin{equation}
|{\cal M_{\alpha \beta}}|^2
=
|{\cal M}_{\rm SM,{\alpha \beta}}|^2
+
|{\cal M}_{H,{\alpha \beta}}|^2
+
2{\rm Re}
\left(
{\cal M}_{\rm SM,{\alpha \beta}}{\cal M}_{H,{\alpha \beta}}^{*}
\right),
\end{equation}
where the last term denotes the interference between the heavy Higgs mediated amplitude and the continuum SM contribution. 
For the CP-even scalar $H$, its contribution enters through the $s$-channel resonance, and the corresponding amplitude is given by
\begin{equation}
{\cal M}_{H,{\alpha \beta}} = {\cal V}_{\gamma\gamma H} \frac{1} {\hat{s}-m_H^2+i m_H\Gamma_H} {\cal V}_{Htt}.
\end{equation}
Here, ${\cal V}_{\gamma\gamma H}$ and ${\cal V}_{Htt}$ denote the effective vertex factors for the $H\gamma\gamma$ and $Ht\bar t$ interactions, respectively, while $m_H$ and $\Gamma_H$ are the mass and total decay width of the heavy Higgs boson, respectively.

Since the coupling between a CP-even Higgs boson and two photons is absent at tree level, the $H\gamma\gamma$ vertex is generated by charged-particle loop contributions. The corresponding effective interaction can be
described by
\begin{equation}
{\cal L}_{H\gamma\gamma}
=
-\frac14 g_{H\gamma\gamma}
H F_{\mu\nu}F^{\mu\nu}.
\end{equation}
The effective coupling $g_{H\gamma\gamma}$ is proportional to the loop-induced form factor ${\cal F}_{H\gamma\gamma}$,
\begin{equation}
{\cal F}_{H\gamma\gamma}
=
\kappa_W^H A_1(\tau_W)
+
\sum_f N_c^f Q_f^2 \kappa_f^H A_{1/2}(\tau_f)
+
\kappa_{H^\pm}A_0(\tau_{H^\pm}),
\end{equation}
Here
$\tau_i=\frac{4m_i^2}{\hat{s}}$,
$N_c^f$ and $Q_f$ denote the color factor and electric charge of the fermion $f$, respectively, while the coupling modifiers $\kappa_f^H$ and $\kappa_{H^\pm}$, as well as the loop functions $A_1$, $A_{1/2}$, and $A_0$, follow the conventions of Ref.~\cite{Xu:2025vmy}.

The Yukawa interaction between the heavy Higgs boson and top quarks is given by~\cite{Aoki:2009ha}
\begin{equation}
{\cal L}_{Htt}
=
-\frac{m_t}{v}
\kappa_t^H H\bar t t ,
\end{equation}
where the coupling modifier in the Type-II 2HDM is
\begin{equation}
\kappa_t^H
=
\cos{(\beta-\alpha)}
-
\sin{(\beta-\alpha)}\cot\beta.
\end{equation}
We adopt the alignment limit, $\sin{(\beta-\alpha)}=1$, in which the lighter CP-even Higgs boson $h$ has SM-like couplings, and the above coupling reduces to $\kappa_t^H=-\cot\beta$~\cite{Branco:2011iw,Bernon:2015wef}.
This coupling, together with the effective $H\gamma\gamma$ vertex, controls the heavy Higgs contribution to the $\gamma\gamma\rightarrow t\bar t$ process.
In the spin density matrix formalism, the production of the $t\bar t$ system in $\gamma\gamma$ collisions is described by the unnormalized production spin density matrix $R$.
In the spin basis of the top and anti-top quarks, its matrix elements are given by
\begin{equation}
R_{\alpha\beta,\alpha'\beta'}
=
\mathcal{M}_{\alpha\beta}
\mathcal{M}_{\alpha'\beta'}^*.
\end{equation}
The density matrix can then be decomposed in the two-particle spin space as~\cite{Bernreuther:2015yna}
\begin{equation}
R
=
\tilde A\,\mathbb{1}\otimes\mathbb{1}
+
\tilde B_i^{+}\,\sigma^i\otimes\mathbb{1}
+
\tilde B_i^{-}\,\mathbb{1}\otimes\sigma^i
+
\tilde C_{ij}\,\sigma^i\otimes\sigma^j .
\end{equation}
Here, $\mathbb{1}$ denotes the $2\times 2$ identity matrix and $\sigma^{i(j)}$ ($i,j=1,2,3$) are the Pauli matrices acting in the two-dimensional spin space of the top or anti-top quark.
The scalar coefficient $\tilde A$ denotes the unpolarized production rate, while the vectors $\tilde B_i^{+}$ and $\tilde B_i^{-}$ describe the single-particle polarizations of the top and anti-top quarks, respectively, and the tensor $\tilde C_{ij}$ represents their spin correlations.

To analyze the kinematic structure of these coefficients, we work in the $t\bar t$ zero-momentum frame (ZMF) and introduce the right-handed orthonormal basis
$\{\hat{\mathbf r},\hat{\mathbf k},\hat{\mathbf n}\}$, where $\hat{\mathbf k}$ is the direction of the top quark and $\hat{\mathbf p}$ denotes the incoming photon direction~\cite{Bernreuther:2015yna}. 
The remaining basis vectors are defined as
\begin{equation}
\hat{\mathbf r}=\frac{1}{\sqrt{1-z^2}}(\hat{\mathbf p}-z\hat{\mathbf k}),
\qquad
\hat{\mathbf n}=\frac{1}{\sqrt{1-z^2}}(\hat{\mathbf p}\times\hat{\mathbf k}),
\end{equation}
with
$z=\hat{\mathbf p}\cdot\hat{\mathbf k}$.
In this basis, the polarization vectors and spin-correlation tensor can be expanded as
\begin{align}
\tilde B_i^\pm &= b_r^\pm \hat r_i + b_k^\pm \hat k_i + b_n^\pm \hat n_i ,\\
\tilde C_{ij} &=
c_{rr}\hat r_i\hat r_j + c_{kk}\hat k_i\hat k_j + c_{nn}\hat n_i\hat n_j
+ c_{rk}\hat r_i\hat k_j+c_{kr}\hat k_i\hat r_j
+ c_{rn}\hat r_i\hat n_j+c_{nr}\hat n_i\hat r_j
+ c_{kn}\hat k_i\hat n_j+c_{nk}\hat n_i\hat k_j.
\end{align}
The coefficients $b_v^\pm$ and $c_{vv'}$ are scalar functions of the partonic center-of-mass energy $\hat s$ and the scattering-angle variable $z$.

For the construction of physical observables, it is convenient to introduce the normalized density matrix
\begin{equation}
\rho = \frac{R}{\mathrm{Tr}[R]}, \qquad \mathrm{Tr}[\rho]=1.
\end{equation}
In terms of the normalized polarization vectors and correlation tensor,
the density matrix can be written as
\begin{equation}
\rho =
\frac14
\left[
\mathbb{1}\otimes\mathbb{1}
+
B_i^+\,\sigma^i\otimes\mathbb{1}
+
B_i^-\,\mathbb{1}\otimes\sigma^i
+
C_{ij}\,\sigma^i\otimes\sigma^j
\right].
\end{equation}
The normalized polarization vectors and correlation tensor are defined as
\begin{equation}
B_i^\pm = \frac{\tilde B_i^\pm}{\tilde A}, 
\qquad
C_{ij} = \frac{\tilde C_{ij}}{\tilde A}.
\end{equation}

The single-particle polarization along an arbitrary spin quantization axis $\hat{\mathbf a}$ is defined as
\begin{equation}
B_a^\pm = 2\left\langle \hat{\mathbf a}\cdot \vec s_{t,\bar t} \right\rangle ,
\end{equation}
where $\vec s_{t,\bar t}$ denotes the spin operator of the top (anti-top) quark. 
Similarly, the spin-correlation coefficients are defined by
\begin{equation}
C_{ab}
=
4\left\langle (\hat{\mathbf a}\cdot \vec s_t)(\hat{\mathbf b}\cdot \vec s_{\bar t}) \right\rangle ,
\end{equation}
with $\hat{\mathbf a}$ and $\hat{\mathbf b}$ being the chosen spin-projection axes for the top and anti-top quarks, respectively. 
Equivalently, these observables correspond to projections of the polarization vectors and correlation tensor,
\begin{equation}
B_a^\pm = a_i B_i^\pm,
\qquad
C_{ab} = a_i b_j C_{ij}.
\end{equation}
The quantities $B_i^\pm$ and $C_{ij}$ therefore encode the complete polarization and spin-correlation information of the normalized two-spin state, while the measurable observables $B_a^\pm$ and $C_{ab}$ are obtained by projection onto specific spin quantization axes.

As a quantitative measure of entanglement, we employ the concurrence~\cite{Wootters:1997id}
\begin{equation}
\mathcal{C}[\rho] =
\max\left(0, r_1-r_2-r_3-r_4\right),
\end{equation}
where $r_1 \ge r_2 \ge r_3 \ge r_4$, and  $r_i$ are the square roots of the eigenvalues of
\begin{equation}
\mathcal{R}=
\rho(\sigma_2\otimes\sigma_2)\rho^*(\sigma_2\otimes\sigma_2),
\end{equation}
with $\rho^*$ denoting complex conjugation and $\sigma_2$ the second Pauli matrix.

Nonlocal correlations are tested via the CHSH inequality~\cite{Clauser:1969ny, Horodecki:1995nsk}. 
Defining the matrix
\begin{equation}
M = C C^T,
\end{equation}
where $C$ denotes the $3\times3$ correlation matrix formed by the coefficients $C_{ij}$, and letting
$m_1 \ge m_2 \ge m_3$
be the eigenvalues of $M$, the CHSH inequality is violated when
\begin{equation}
\mathfrak{B} \equiv m_1 + m_2 > 1.
\end{equation}

\section{ Numerical results and analysis}\label{sec4}
In this section, we present the numerical results for the process $\gamma\gamma\to t\bar t$ in the Type-II 2HDM at LO. 
We investigate the total cross section, single-particle polarization and spin-correlation observables, as well as the quantum entanglement properties of the $t\bar t$ system. 
Particular attention is paid to the effects of the heavy CP-even Higgs boson $H$ and its interference with the SM contribution.

\subsection{Selection of Input Parameters}
The input parameters for SM adopted in our calculations are specified as follows~\cite{ParticleDataGroup:2024cfk}:
\begin{equation}
\begin{aligned}
\label{9}
&  ~~~~~~ m_W=80.369\ \mathrm{GeV},   ~~~~~~ m_b=4.18 \ \mathrm{GeV}, ~~~~~~ m_h = 125.20\ \mathrm{GeV},\\
&  ~~~~~~ m_t=172.4 \ \mathrm{GeV},    ~~~~~~ m_e=0.511 \ \mathrm{MeV}, ~~~~~~ \alpha^{-1} = 128, ~~~~~~ \alpha_s(\mu=m_t) = 0.1 .
\end{aligned}
\end{equation}

In the numerical analysis, we consider the Type-II 2HDM and focus on the heavy CP-even Higgs boson \(H\). The parameter space is required to satisfy the theoretical constraints from vacuum stability, perturbativity, and unitarity, together with the electroweak precision constraints implemented in \texttt{2HDMC}~\cite{Eriksson:2009ws}. To satisfy the constraints from the oblique parameter \(T\)~\cite{Peskin:1991sw} and flavor observables~\cite{Atkinson:2021eox}, we take $m_A=m_{H^\pm}=900~\mathrm{GeV}$.
Under the alignment limit, the heavy Higgs contribution is governed by $m_H$, $\tan\beta$, and $m_{12}^2$. 
The lower and upper bounds of the $m_H$ scan are chosen near the $t\bar t$ production threshold and at the maximum center-of-mass energy of the photon-photon system, respectively, for the adopted Compton backscattering setup. 
Given the stringent constraints from existing heavy neutral Higgs searches on the low-$\tan\beta$ region~\cite{ATLAS:2024vxm,CMS:2025dzq}, we restrict the scan to $\tan\beta\geq 3$.
 The parameter ranges are
\begin{equation*}
350~\mathrm{GeV}\leq m_H \leq 828.4~\mathrm{GeV},\qquad
3\leq\tan\beta\leq10,\qquad
1\times10^4~\mathrm{GeV}^2<m_{12}^{2}<4\times10^5~\mathrm{GeV}^2.
\end{equation*}

\begin{figure}[!htbp]
	\centering
	\includegraphics[width=0.7\linewidth]{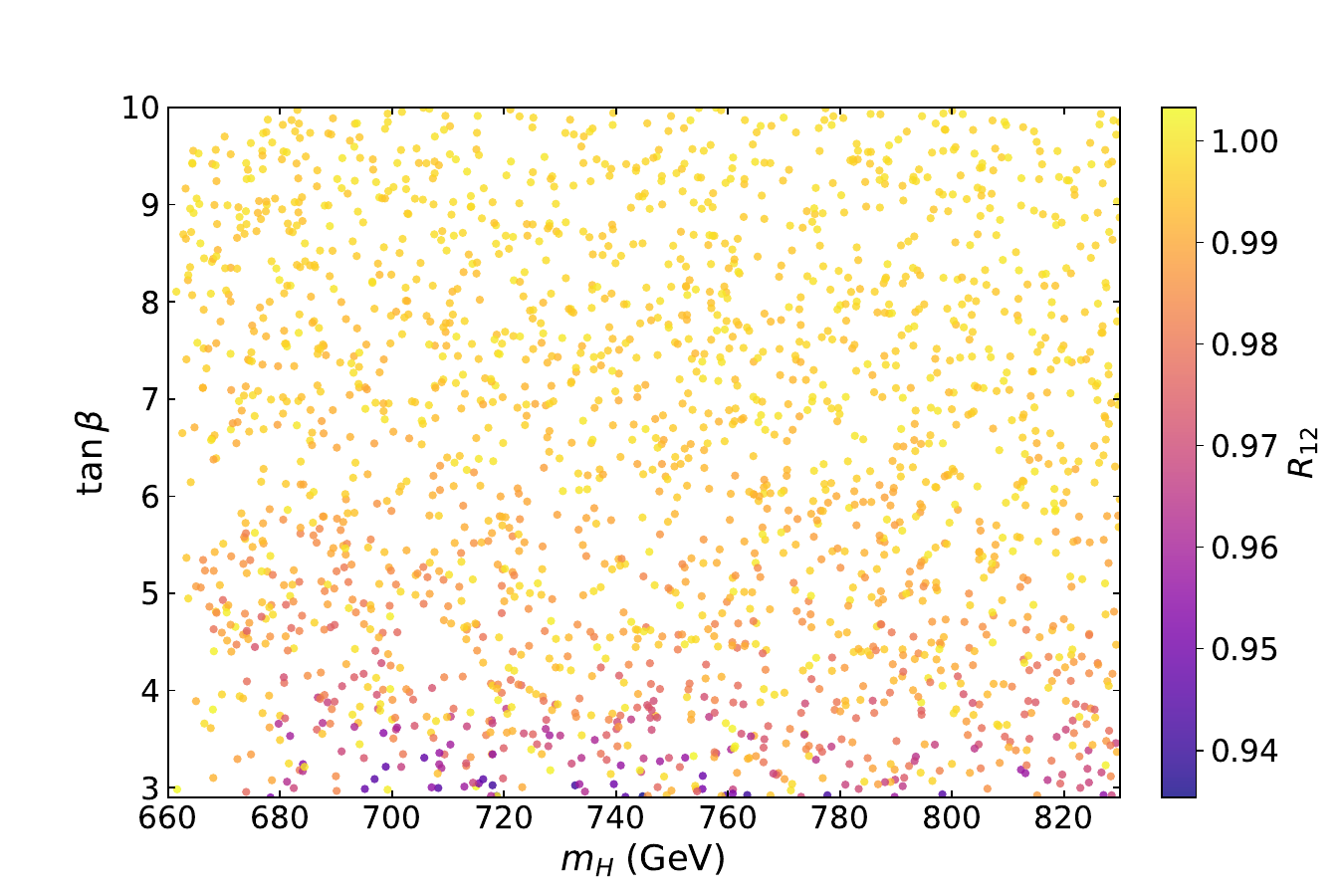}
	\caption{Viable parameter points in the \(m_H\)--\(\tan\beta\) plane after imposing the theoretical and electroweak precision constraints in the Type-II 2HDM, with \(m_A=m_{H^\pm}=900~\mathrm{GeV}\). The color indicates \(m_{12}^{2}/(m_H^{2}\sin \beta \cos \beta)\).}
	\label{fig:scan_2hdm}
\end{figure}

Figure~\ref{fig:scan_2hdm} shows the viable parameter points in the \(m_H\)--\(\tan\beta\) plane, with the color indicating $R_{12} \equiv m_{12}^{2} / \left(m_{H}^{2}\sin\beta\cos\beta\right)$.
With $m_A=m_{H^\pm}=900~\mathrm{GeV}$ fixed, the increasing scalar mass splittings at lower $m_H$ lead to more restrictive perturbativity constraints, leaving few viable points below $660~\mathrm{GeV}$, so we focus on $660\leq m_H\leq828.4~\mathrm{GeV}$.
Since the viable points are predominantly clustered around \(R_{12}\sim1\), we adopt \(m_{12}^{2}=m_H^{2}\sin\beta\cos\beta\) in the subsequent analysis.

\begin{figure}[!htbp]
	\begin{center}
		\includegraphics[width=0.9\textwidth]{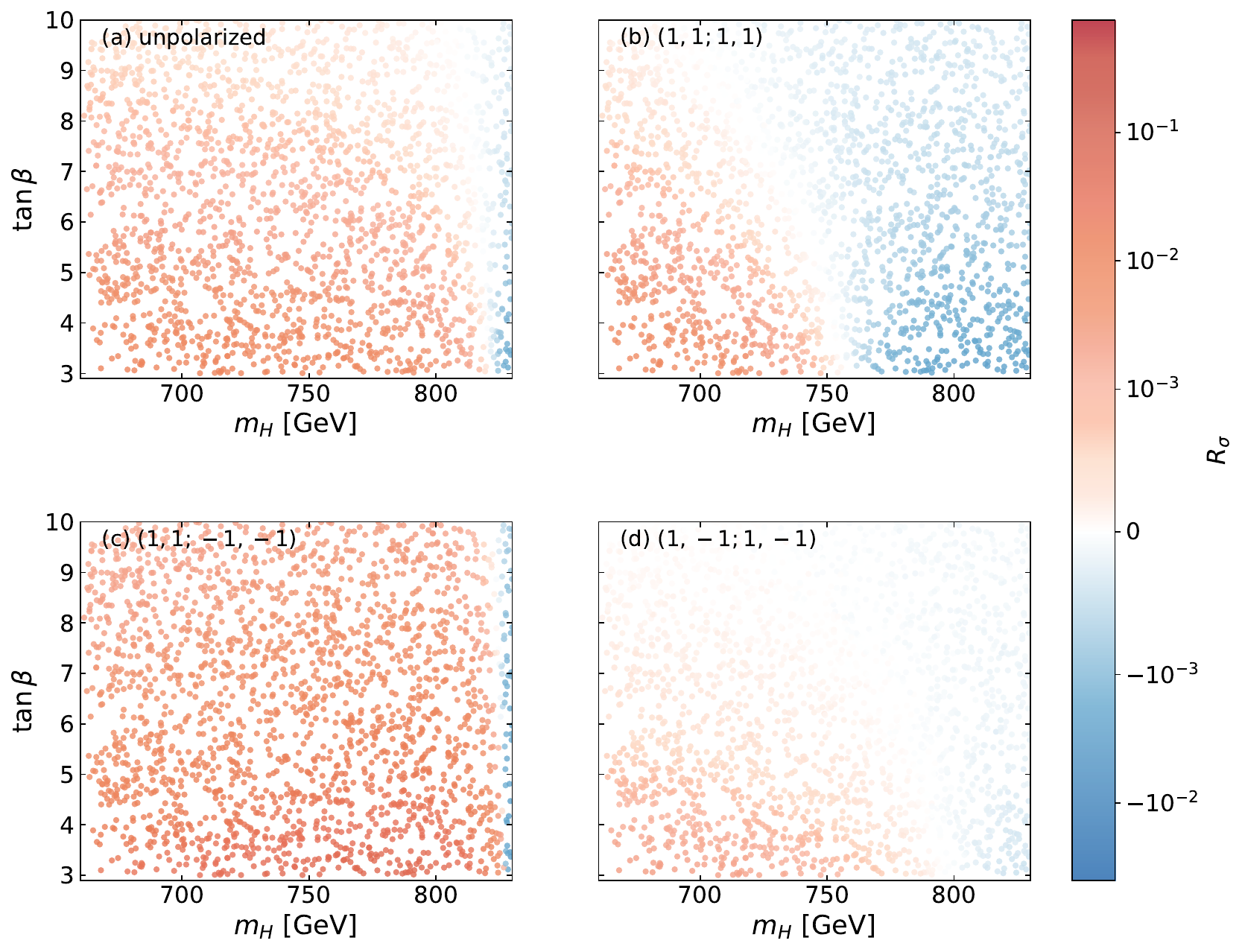}
        \caption{
         Relative deviation \(R_{\sigma}=(\sigma_{\rm Total}-\sigma_{\rm SM})/\sigma_{\rm SM}\) in the \(m_H\)--\(\tan\beta\) plane from the Type-II 2HDM parameter scan. Here, \(\sigma_{\rm Total}\) includes the heavy-Higgs contribution and its interference with the SM amplitude, while \(\sigma_{\rm SM}\) denotes the SM cross section. Panels (a)--(d) correspond to the unpolarized case and the initial-state beam polarization configurations \((P_e^{(1)},P_e^{(2)};P_L^{(1)},P_L^{(2)})=(1,1,1,1)\), \((1,1,-1,-1)\), and \((1,-1,1,-1)\), respectively.
}
		\label{fig:scan}
	\end{center}
\end{figure}

Building on the viable parameter space obtained above, we further investigate the relative deviation
\(R_{\sigma}=(\sigma_{\rm Total}-\sigma_{\rm SM})/\sigma_{\rm SM}\) 
in the \(m_H\)--\(\tan\beta\) plane for different initial-state photon polarization configurations, where \(\sigma_{\rm Total}\) denotes the cross section including both the SM and heavy-Higgs-mediated amplitudes and their interference, while \(\sigma_{\rm SM}\) is the corresponding SM prediction.
We consider the unpolarized case as well as the configurations \((1,1;1,1)\), \((1,1;-1,-1)\), and \((1,-1;1,-1)\), where the four entries denote $(P_e^{(1)}, P_e^{(2)} ; P_L^{(1)}, P_L^{(2)}) $, respectively, as shown in Fig.~\ref{fig:scan}. 
The red and blue regions correspond to \(R_{\sigma}>0\) and \(R_{\sigma}<0\), respectively, with the color intensity indicating the magnitude of \(|R_{\sigma}|\), while the white regions correspond to \(R_{\sigma}\simeq0\).

We find that the magnitude of \(R_{\sigma}\) generally decreases with increasing \(\tan\beta\), with larger deviations occurring predominantly in the relatively low-\(\tan\beta\) region of the scanned parameter space. 
For the unpolarized, \((1,1;1,1)\), and \((1,-1;1,-1)\) configurations, the regions with sizable positive \(R_{\sigma}\) are mainly concentrated around \(m_H\sim670\)--\(700~\mathrm{GeV}\). 
Among these configurations, the \((1,-1;1,-1)\) configuration gives the weakest effect. 
In contrast, the \((1,1;-1,-1)\) configuration exhibits a different dependence on the model parameters, with sizable positive deviations extending to a higher-mass region of \(m_H\sim750\)--\(780~\mathrm{GeV}\). 
These differences originate from the distinct weights with which the initial-state photon helicity configurations contribute to the heavy-Higgs resonance and its interference with the SM amplitude, resulting in different parameter dependences of \(R_{\sigma}\).

Taking into account the experimental constraints on heavy neutral Higgs bosons, we select
$m_H=678.55~\mathrm{GeV},
\tan\beta=3.08$ and 
$m_{12}^2=1.35\times10^5~\mathrm{GeV}^2$ 
as the benchmark point for the subsequent analysis. 
This point satisfies the theoretical and electroweak precision constraints, and its mass spectrum with $m_H< m_A=m_{H^\pm}=900~\mathrm{GeV}$ is consistent with the theoretically allowed configurations discussed in Ref.~\cite{Kling:2018xud}. 
Moreover, the mass of $H$ is sufficiently separated from the fixed masses of $A$ and $H^\pm$. 
We use this benchmark to examine the impact of the heavy Higgs boson on top-quark pair spin correlations and quantum entanglement observables.

\subsection{Cross Sections}
At the benchmark point, the cross sections for different initial-state polarization configurations are summarized in Table~\ref{tab:sigma}.
The ratio $\mu_H=\sigma_{\mathrm{Total}}/\sigma_{\mathrm{SM}}$ quantifies the effect of the heavy Higgs boson on the inclusive cross section. 
It ranges from 1.00 to 1.02 for the considered polarization configurations.
The largest enhancement is obtained for the $(1,1;-1,-1)$ configuration, where the cross section increases by about $2.3\%$ relative to the SM prediction, while the effect is much smaller for the $(1,-1;1,-1)$ configuration.
This dependence reflects the different contributions of the polarized photon initial states to the SM and heavy-Higgs-mediated amplitudes, as well as their interference.

\begin{table}[ht]
	\centering
	\setlength{\tabcolsep}{7pt}
	\renewcommand{\arraystretch}{1.0}
	\begin{tabular}{ccccc}
		\hline
		\hline
		&
		Unpolarized
		&
		$(1,1;1,1)$
		&
		$(1,1;-1,-1)$
		&
		$(1,-1;1,-1)$
		
		\\
		\hline
		$\sigma_{\mathrm{Total}}$ [fb]
		& 514.33
		& 602.51
		& 461.51
		& 372.24
		\\
		
		$\sigma_{\mathrm{SM}}$ [fb]
		& 509.07
		& 597.68
		& 451.03
		& 371.35
		\\
		
		$\mu_H=	\sigma_{\mathrm{Total}}/\sigma_{\mathrm{SM}}$
		& 1.01
		& 1.01
		& 1.02
		& 1.00
		\\
		\hline
	\end{tabular}
	\caption{SM and total cross sections and corresponding $\mu_H$-factors for different beam polarization configurations $(P_e^{(1)}, P_e^{(2)} ; P_L^{(1)}, P_L^{(2)}) $ at $\sqrt{s_{ee}}=1000~\mathrm{GeV}$.}
	\label{tab:sigma}
\end{table}


\begin{figure}[!htbp]
	\begin{center}
			\includegraphics[width=0.8\textwidth]{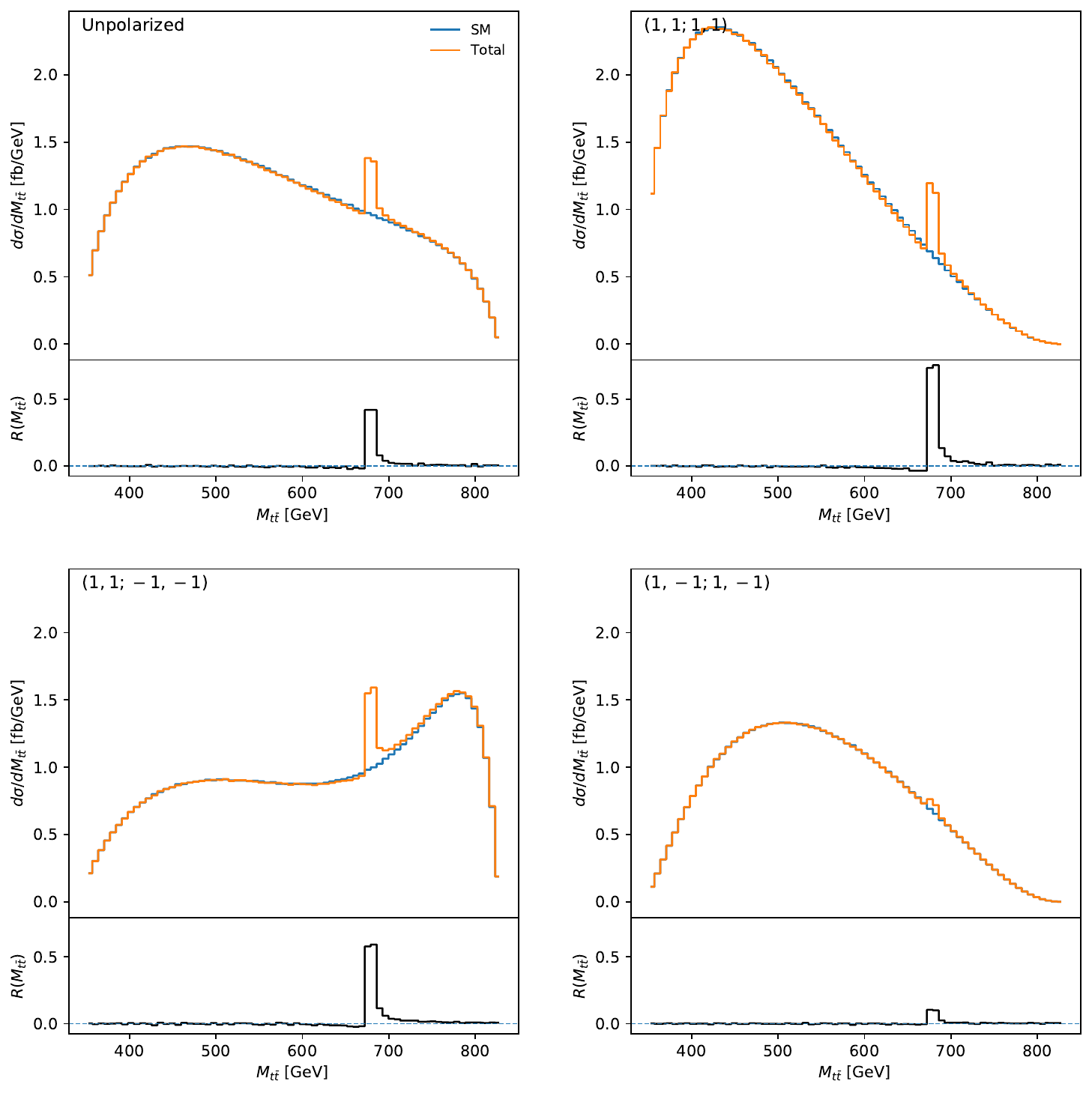}
\caption{
Invariant mass distribution of the \(t\bar{t}\) system for the SM and the full prediction including the heavy CP-even Higgs contribution in the type-II 2HDM, for different initial-state photon polarization configurations. The upper panel shows \(d\sigma/dM_{t\bar{t}}\), while the lower panel displays the relative differential variation \(R(M_{t\bar{t}})\).
}
	\label{fig:Mtt_distribution}
	\end{center}
\end{figure}

Figure~\ref{fig:Mtt_distribution} shows the invariant mass distributions of the top-quark pair for different photon polarization configurations, comparing the SM prediction with the total result including the heavy Higgs boson.
The two distributions are close to each other away from the resonance region, while a visible deviation appears around $M_{t\bar t}\simeq m_H$.
The corresponding relative modification,
$$R(M_{t\bar t})=
\frac{
d\sigma_{\rm Total}/dM_{t\bar t}
-
d\sigma_{\rm SM}/dM_{t\bar t}
}{
d\sigma_{\rm SM}/dM_{t\bar t}
}$$
is shown in the lower panels.
It quantifies the difference between the two distributions and reaches its largest values around the heavy Higgs resonance.
For the $(1,1;1,1)$ configuration, $R(M_{t\bar t})$ exceeds approximately $0.75$, while the corresponding modification is much smaller for $(1,-1;1,-1)$.
This localized enhancement is much more pronounced than the corresponding modification of the inclusive cross section.

\subsection{ Polarization and Spin correlation Observables}

Table~\ref{tab:Ba} presents the single-particle polarization coefficients for the different photon-polarization configurations at $\sqrt{s_{ee}}=1000$~GeV.
In the chosen basis and for the initial-state configurations considered here, $B_n^\pm$ vanishes and is therefore not shown.
In addition, the top and antitop polarizations satisfy $B_k^-=-B_k^+$ and $B_r^-=B_r^+$, so only $B_k^+$ and $B_r^+$ are listed.
For the $(1,1;1,1)$ and $(1,1;-1,-1)$ configurations, the polarization is along $\hat{\mathbf k}$, while for $(1,-1;1,-1)$, it is along $\hat{\mathbf r}$.
Including the heavy Higgs contribution changes these coefficients only slightly, with the ratios $R_B=B_i^{\rm Total}/B_i^{\rm SM}$ remaining within about $1\%$ of unity for the nonvanishing polarization components.
The polarization pattern is therefore unchanged, with the heavy-Higgs contribution producing only a small modification of the single-particle polarization magnitude.
\begin{table}[!hbp]
	\centering
	\setlength{\tabcolsep}{8pt}
	\renewcommand{\arraystretch}{1.0}
	\begin{tabular}{c|ccc|ccc|ccc|ccc}
		\hline
		\hline
		& \multicolumn{3}{c|}{Unpolarized}
		& \multicolumn{3}{c|}{$(1,1;1,1)$}
		& \multicolumn{3}{c|}{$(1,1;-1,-1)$}
		& \multicolumn{3}{c}{$(1,-1;1,-1)$}
		\\
		& Total & SM & $R_B$
		& Total & SM & $R_B$
		& Total & SM & $R_B$
		& Total & SM & $R_B$
		\\
		\hline
		$B_k^+$
		& 0 & 0 & --
		& 0.844 & 0.850 & 0.993
		& 0.433 & 0.429 & 1.009
		& 0 & 0 & --
		\\
		$B_r^+$
		& 0 & 0 & --
		& 0 & 0 & --
		& 0 & 0 & --
		& 0.595 & 0.596 & 0.998
		\\
		\hline
	\end{tabular}
	\caption{Single-particle polarization coefficients for the total and SM contributions at $\sqrt{s_{ee}}=1000$ GeV under various polarized photon configurations. The ratio is defined as $R_B=B_i^{\rm Total}/B_i^{\rm SM}$ and is not defined when the corresponding SM coefficient vanishes.}
	\label{tab:Ba}
\end{table}

Table~\ref{tab:Cab} presents the diagonal spin-correlation coefficients for different photon-polarization configurations at $\sqrt{s_{ee}}=1000$~GeV. 
Compared with the percent-level changes in the inclusive cross section, the heavy Higgs contribution induces more pronounced modifications in some spin-correlation coefficients, with a strong dependence on the initial-state photon polarization. 
For the unpolarized configuration, all three diagonal coefficients receive some modifications. For the $(1,1;1,1)$ configuration, $C_{kk}$ remains essentially unchanged, while $C_{rr}$ and $C_{nn}$ are modified to some extent. 
The changes in all three diagonal coefficients are more pronounced for the $(1,1;-1,-1)$ configuration, whereas the diagonal correlations remain nearly unchanged for $(1,-1;1,-1)$. 
These results indicate that the impact of the heavy Higgs contribution on the $t\bar{t}$ spin correlations is concentrated in specific initial-state polarization configurations and spin directions.

For the off-diagonal spin correlations, $C_{kr}$, $C_{rk}$, $C_{nk}$ and $C_{kn}$ vanish for all the polarization configurations considered. 
The coefficients $C_{rn}$ and $C_{nr}$ exhibit a different behavior: they vanish for all configurations in the SM, while the inclusion of the heavy Higgs contribution gives $C_{rn}=-0.010$ and $0.007$ for the $(1,1;1,1)$ and $(1,1;-1,-1)$ configurations, respectively, with $C_{nr}=-C_{rn}$. 
They remain zero for the unpolarized and $(1,-1;1,-1)$ configurations. 
This shows that the heavy Higgs contribution not only modifies the magnitude of selected diagonal spin correlations, but can also induce off-diagonal spin correlations absent in the SM for specific initial-state photon-polarization configurations.
\begin{table}[!htbp]
	\centering
	\setlength{\tabcolsep}{6pt}
	\renewcommand{\arraystretch}{1.0}
	\begin{tabular}{c|ccc|ccc|ccc|ccc}
		\hline
		\hline
		& \multicolumn{3}{c|}{Unpolarized}
		& \multicolumn{3}{c|}{$(1,1;1,1)$}
		& \multicolumn{3}{c|}{$(1,1;-1,-1)$}
		& \multicolumn{3}{c}{$(1,-1;1,-1)$}
		\\
		& Total & SM & $R_C$
		& Total & SM & $R_C$
		& Total & SM & $R_C$
		& Total & SM & $R_C$
		\\
		\hline
		$C_{kk}$
		& -0.343 & -0.336 & 1.021
		& -0.922 & -0.922 & 1.000
		& -0.344 & -0.328 & 1.049
		& 0.432 & 0.436 & 0.991
		\\
		$C_{rr}$
		& 0.052 & 0.042 & 1.238
		& -0.326 & -0.337 & 0.967
		& 0.177 & 0.158 & 1.120
		& 0.539 & 0.538 & 1.002
		\\
		$C_{nn}$
		& -0.299 & -0.312 & 0.958
		& -0.367 & -0.378 & 0.971
		& -0.187 & -0.215 & 0.870
		& -0.252 & -0.255 & 0.988
		\\
		\hline
	\end{tabular}
	\caption{Diagonal spin-correlation coefficients for the total and SM contributions
	at $\sqrt{s_{ee}}=1000$ GeV under unpolarized and various polarized photon
	configurations. The ratio is defined as $R_C=C_{ij}^{\rm Total}/C_{ij}^{\rm SM}$.}
	\label{tab:Cab}
\end{table}

\subsection{Entanglement Observables}
Building upon our previous study of quantum entanglement in \(\gamma\gamma\to t\bar t\) with NLO QCD corrections~\cite{Jiang:2026sfw}, we extend the analysis to include the heavy CP-even Higgs contribution in the Type-II 2HDM, presenting SM predictions at LO and NLO QCD and 2HDM results at LO.
Since the isolated $H$-mediated amplitude produces a maximally entangled $t\bar t$ spin state, with $\mathcal C[\rho]=1$, it is particularly interesting to examine how the inclusion of the SM amplitude and its interference with the Higgs amplitude affect the entanglement properties of the resulting $t\bar t$ state.

\begin{figure}[!htbp]
	\begin{center}
			\includegraphics[width=0.8\textwidth]{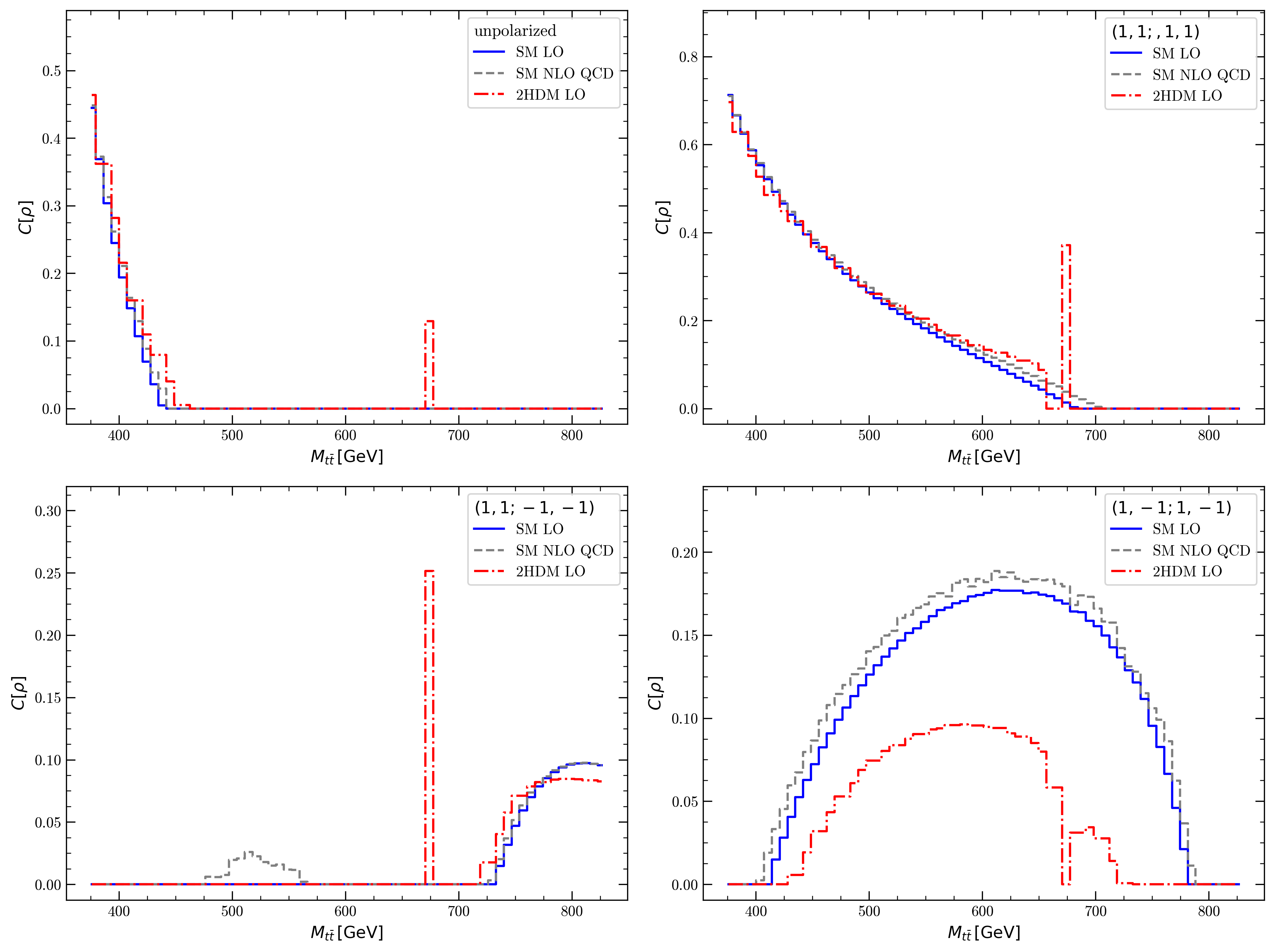} 	
\caption{Invariant-mass distributions of the concurrence for the unpolarized, $(1,1;1,1)$, $(1,1;-1,-1)$, and $(1,-1;1,-1)$ photon-polarization configurations, shown in the four panels, respectively.
The blue solid, gray dotted, and red dashed curves represent the SM LO, SM NLO QCD, and 2HDM LO predictions, respectively.}
	\label{fig:C_4panel}
	\end{center}
\end{figure}

Figure~\ref{fig:C_4panel} shows the invariant-mass distributions of the concurrence for different initial-state photon polarizations.
For the unpolarized and $(1,1;1,1)$ configurations, the SM concurrence decreases with increasing $M_{t\bar t}$ and approaches zero in the high-invariant-mass region.
The NLO QCD corrections lead to only modest changes in the SM LO predictions, whereas the heavy Higgs contribution induces more pronounced modifications, particularly an enhancement around the resonance, together with sizable deviations from the SM results away from the resonance.
For the $(1,1;-1,-1)$ configuration, the SM LO concurrence is nearly zero over the low and intermediate mass regions, while the NLO QCD corrections give rise to a nonzero contribution around $M_{t\bar t}\sim 500$ GeV.
By contrast, the heavy Higgs contribution generates a pronounced entanglement structure around the resonance, with the concurrence reaching approximately $0.25$.
Interestingly, all three predictions exhibit a further enhancement of the concurrence in the high-invariant-mass region.
This behavior can be understood from the increasingly pronounced longitudinal spin anti-correlation between the top and antitop, accompanied by comparatively weaker transverse spin correlations. 
For the $(1,-1;1,-1)$ configuration, the SM concurrence first increases and then decreases with $M_{t\bar t}$, with only a small modification from the NLO QCD corrections.
The heavy Higgs contribution, however, suppresses the concurrence over a broad invariant-mass range and drives it to zero around the resonance.

\begin{figure}[!htbp]
	\begin{center}
			\includegraphics[width=0.8\textwidth]{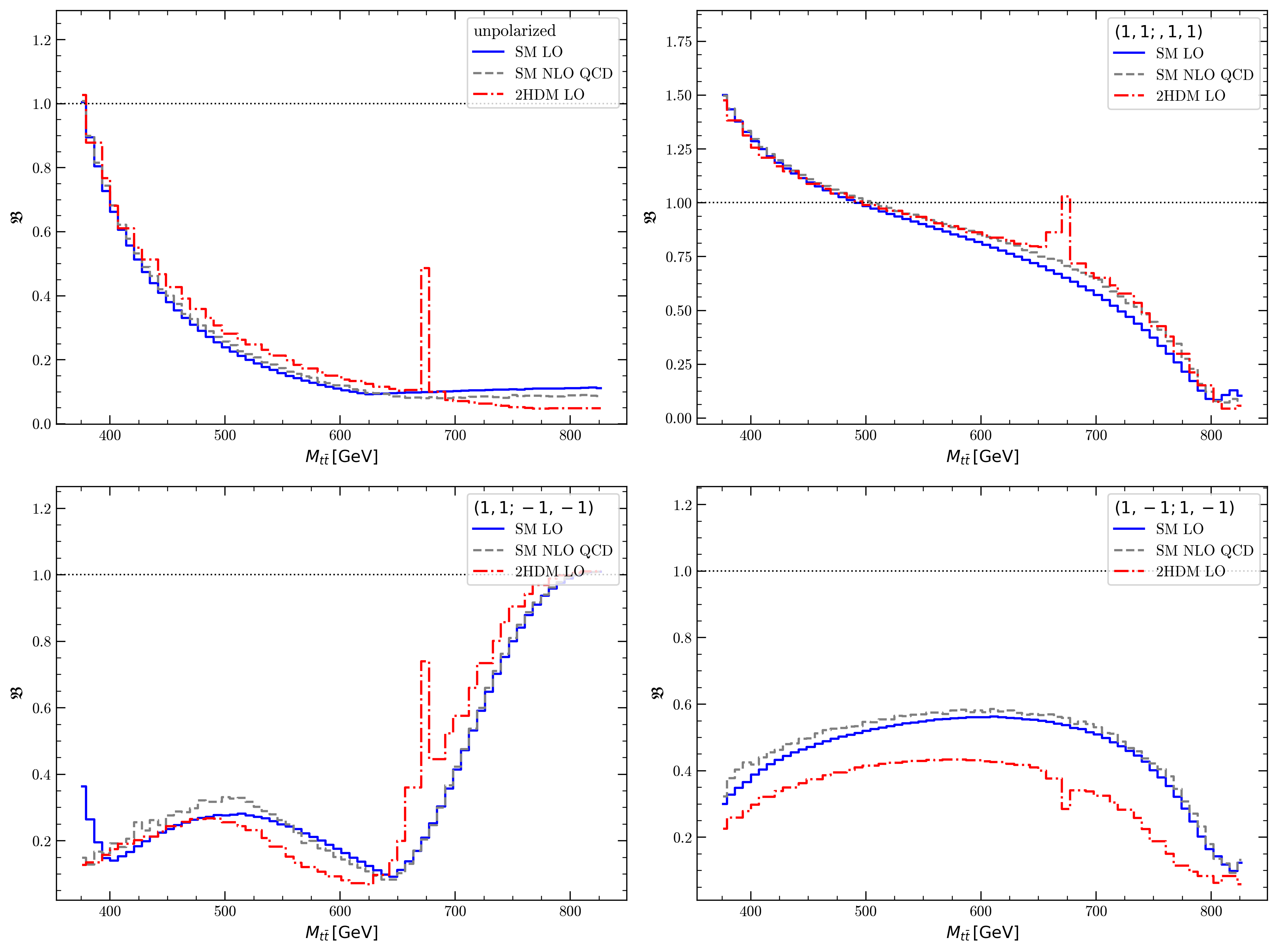} 	
\caption{Invariant-mass distributions of the Bell--CHSH observable $\mathfrak{B}$ for the unpolarized, $(1,1;1,1)$, $(1,1;-1,-1)$, and $(1,-1;1,-1)$ photon-polarization configurations, shown in the four panels, respectively.
The blue solid, gray dotted, and red dashed curves represent the SM LO, SM NLO QCD, and 2HDM LO predictions, respectively.
The horizontal line at $\mathfrak{B}=1$ indicates the threshold for Bell--CHSH inequality violation.}
	\label{fig:m12_4panel}
	\end{center}
\end{figure}

Figure~\ref{fig:m12_4panel} shows the dependence of the Bell parameter \(\mathfrak{B}\) on the invariant mass. 
While the isolated $H$-mediated contribution yields a maximally entangled \(t\bar t\) spin state, the full 2HDM LO prediction also includes the SM contribution and its interference with the heavy-Higgs amplitude, resulting in deviations from maximal entanglement.
For unpolarized photons, all three predictions exceed unity only slightly near the $t\bar t$ production threshold and decrease rapidly with increasing $M_{t\bar t}$.
For the $(1,1;1,1)$ configuration, both the SM LO and NLO QCD predictions decrease from values above unity near threshold and eventually fall below the Bell-CHSH threshold.
The 2HDM prediction also decreases below unity in the low-invariant-mass region, but exhibits a pronounced enhancement around the heavy Higgs resonance, reaching approximately $1.02$ and thus exceeding the Bell-CHSH threshold in this region.
For the $(1,1;-1,-1)$ configuration, $\mathfrak{B}$ remains below unity over most of the considered invariant-mass range and exceeds the threshold only at high invariant masses.
The heavy Higgs contribution slightly extends the mass region in which this condition is satisfied.
This behavior is consistent with the enhancement of the concurrence in the same high-invariant-mass region, where the increasingly pronounced longitudinal spin anticorrelation contributes to the entanglement of the $t\bar t$ pair.
In contrast, for the $(1,-1;1,-1)$ configuration, all three predictions remain below unity throughout the considered mass range.
These results show that the heavy Higgs contribution can substantially modify the invariant-mass dependence of $\mathfrak{B}$. 
For certain initial-state photon polarizations, it can also alter the kinematic region in which the CHSH Bell inequality is violated, whereas the NLO QCD corrections have a comparatively minor impact on this criterion.

\newpage
\section{Summary}\label{sec5}
We have studied the effects of the heavy CP-even Higgs boson $H$ in the Type-II 2HDM on spin correlations and quantum entanglement in $\gamma\gamma\to t\bar t$. 
After imposing the theoretical, electroweak precision, and experimental constraints, we selected a viable benchmark point and compared the SM predictions at LO with the 2HDM results at LO for different initial-state photon polarization configurations.

The numerical results show that the heavy Higgs contribution has a relatively limited impact on conventional production observables. 
Although a visible enhancement appears in the invariant-mass distribution around the resonance, the maximum variation in the inclusive cross section is only about $2.3\%$. 
The single-particle polarization coefficients are also modified only at the percent level. 
In contrast, the heavy Higgs contribution leads to more pronounced modifications of the $t\bar t$ spin correlations, with a strong dependence on the initial-state photon polarization. 
In addition to modifying selected diagonal spin-correlation coefficients, it induces nonzero off-diagonal correlations that vanish in the SM for specific polarization configurations.

These effects are further reflected in the quantum-entanglement observables. 
The concurrence exhibits distinct behaviors for different photon-polarization configurations. 
For the $(1,1;1,1)$ configuration, the heavy Higgs contribution modifies the entanglement distribution around the resonance and in other invariant-mass regions. 
For the $(1,1;-1,-1)$ configuration, a pronounced entanglement structure emerges around the resonance, with the concurrence reaching approximately $0.25$. 
In contrast, for the $(1,-1;1,-1)$ configuration, the heavy Higgs contribution suppresses the concurrence over a broad invariant-mass range and drives it to zero around the resonance. 
The Bell--CHSH parameter also exhibits a pronounced dependence on the initial-state photon polarization. 
The heavy Higgs contribution modifies its invariant-mass dependence and can alter the kinematic regions in which the Bell--CHSH inequality is violated. 
By comparison, the SM NLO QCD corrections generally induce relatively small changes in these quantum-entanglement observables.

For the benchmark point considered in this work, the heavy-Higgs effects on conventional production observables are modest, whereas more pronounced modifications are found in the spin correlations and quantum-entanglement properties of the $t\bar t$ system. 
This motivates further investigations of spin and quantum-entanglement observables over a broader 2HDM parameter space, which may provide further understanding of the effects of heavy Higgs bosons at future polarized photon colliders.

\section*{Acknowledgments}

The authors thank the members of the theoretical physics group at University of Jinan for helpful discussions and communications. 
We are grateful to Prof. Ke-Ping Xie at the Tsung-Dao Lee Institute, Shanghai Jiao Tong University, for helpful discussions. 
We also thank Prof. Lei Wang at Yantai University for valuable discussions.
This work is supported in part by National Natural Science Foundation of China under the Grants  12321005 and 12235008, Natural Science Foundation of Shandong Province under the Grant ZR2026MS0091 and University of Jinan
Disciplinary Cross-Convergence Construction Project 2024
(XKJC-202404).
\newpage

\bibliography{ref}

@article{Chakraborty:2003iw,
    author = "Chakraborty, Dhiman and Konigsberg, Jacobo and Rainwater, David L.",
    title = "{Review of top quark physics}",
    eprint = "hep-ph/0303092",
    archivePrefix = "arXiv",
    doi = "10.1146/annurev.nucl.53.041002.110601",
    journal = "Ann. Rev. Nucl. Part. Sci.",
    volume = "53",
    pages = "301--351",
    year = "2003"
}

@article{ATLAS:2023fsd,
    author = "Aad, Georges and others",
    collaboration = "ATLAS",
    title = "{Observation of quantum entanglement with top quarks at the ATLAS detector}",
    eprint = "2311.07288",
    archivePrefix = "arXiv",
    primaryClass = "hep-ex",
    reportNumber = "CERN-EP-2023-230",
    doi = "10.1038/s41586-024-07824-z",
    journal = "Nature",
    volume = "633",
    number = "8030",
    pages = "542--547",
    year = "2024"
}

@article{CMS:2024pts,
    author = "Hayrapetyan, Aram and others",
    collaboration = "CMS",
    title = "{Observation of quantum entanglement in top quark pair production in proton{\textendash}proton collisions at $\sqrt{s} = 13$ TeV}",
    eprint = "2406.03976",
    archivePrefix = "arXiv",
    primaryClass = "hep-ex",
    reportNumber = "CMS-TOP-23-001, CERN-EP-2024-137",
    doi = "10.1088/1361-6633/ad7e4d",
    journal = "Rept. Prog. Phys.",
    volume = "87",
    number = "11",
    pages = "117801",
    year = "2024"
}

@article{Bernreuther:2008ju,
    author = "Bernreuther, Werner",
    title = "{Top quark physics at the LHC}",
    eprint = "0805.1333",
    archivePrefix = "arXiv",
    primaryClass = "hep-ph",
    reportNumber = "PITHA-08-09",
    doi = "10.1088/0954-3899/35/8/083001",
    journal = "J. Phys. G",
    volume = "35",
    pages = "083001",
    year = "2008"
}

@article{Mahlon:2010gw,
    author = "Mahlon, Gregory and Parke, Stephen J.",
    title = "{Spin Correlation Effects in Top Quark Pair Production at the LHC}",
    eprint = "1001.3422",
    archivePrefix = "arXiv",
    primaryClass = "hep-ph",
    reportNumber = "FERMILAB-PUB-09-662-T",
    doi = "10.1103/PhysRevD.81.074024",
    journal = "Phys. Rev. D",
    volume = "81",
    pages = "074024",
    year = "2010"
}

@article{Bernreuther:2013aga,
    author = "Bernreuther, Werner and Si, Zong-Guo",
    title = "{Top quark spin correlations and polarization at the LHC: standard model predictions and effects of anomalous top chromo moments}",
    eprint = "1305.2066",
    archivePrefix = "arXiv",
    primaryClass = "hep-ph",
    reportNumber = "TTK-13-13",
    doi = "10.1016/j.physletb.2013.06.051",
    journal = "Phys. Lett. B",
    volume = "725",
    pages = "115--122",
    year = "2013",
    note = "[Erratum: Phys.Lett.B 744, 413--413 (2015)]"
}

@article{CMS:2019nrx,
    author = "Sirunyan, Albert M and others",
    collaboration = "CMS",
    title = "{Measurement of the top quark polarization and $\mathrm{t\bar{t}}$ spin correlations using dilepton final states in proton-proton collisions at $\sqrt{s} =$ 13 TeV}",
    eprint = "1907.03729",
    archivePrefix = "arXiv",
    primaryClass = "hep-ex",
    reportNumber = "CMS-TOP-18-006, CERN-EP-2019-073, CMS-PAS-TOP-18-006",
    doi = "10.1103/PhysRevD.100.072002",
    journal = "Phys. Rev. D",
    volume = "100",
    number = "7",
    pages = "072002",
    year = "2019"
}

@article{ATLAS:2014abv,
    author = "Aad, Georges and others",
    collaboration = "ATLAS",
    title = "{Measurement of Spin Correlation in Top-Antitop Quark Events and Search for Top Squark Pair Production in pp Collisions at $\sqrt{s}=8$ TeV Using the ATLAS Detector}",
    eprint = "1412.4742",
    archivePrefix = "arXiv",
    primaryClass = "hep-ex",
    reportNumber = "CERN-PH-EP-2014-257",
    doi = "10.1103/PhysRevLett.114.142001",
    journal = "Phys. Rev. Lett.",
    volume = "114",
    number = "14",
    pages = "142001",
    year = "2015"
}

@article{Afik:2020onf,
    author = "Afik, Yoav and de Nova, Juan Ram{\'o}n Mu{\~n}oz",
    title = "{Entanglement and quantum tomography with top quarks at the LHC}",
    eprint = "2003.02280",
    archivePrefix = "arXiv",
    primaryClass = "quant-ph",
    doi = "10.1140/epjp/s13360-021-01902-1",
    journal = "Eur. Phys. J. Plus",
    volume = "136",
    number = "9",
    pages = "907",
    year = "2021"
}

@article{Fabbrichesi:2021npl,
    author = "Fabbrichesi, M. and Floreanini, R. and Panizzo, G.",
    title = "{Testing Bell Inequalities at the LHC with Top-Quark Pairs}",
    eprint = "2102.11883",
    archivePrefix = "arXiv",
    primaryClass = "hep-ph",
    doi = "10.1103/PhysRevLett.127.161801",
    journal = "Phys. Rev. Lett.",
    volume = "127",
    number = "16",
    pages = "161801",
    year = "2021"
}

@article{HARLANDER1995137,
title = {Polarization in top quark pair production near threshold},
journal = {Physics Letters B},
volume = {346},
number = {1},
pages = {137-142},
year = {1995},
issn = {0370-2693},
doi = {https://doi.org/10.1016/0370-2693(94)01668-3},
url = {https://www.sciencedirect.com/science/article/pii/0370269394016683},
author = {R Harlander and M Jeżabek and J.H Kühn and T Teubner}
}

@article{Choi:2026omc,
    author = "Choi, Seong Youl and Kang, Dong Woo and Lee, Jae Sik and Park, Chan Beom",
    title = "{Quantum entanglement and Bell nonlocality in top-quark pair production at a photon linear collider}",
    eprint = "2603.12830",
    archivePrefix = "arXiv",
    primaryClass = "hep-ph",
    reportNumber = "IUEP-HEP-26-01",
    month = "3",
    year = "2026"
}

@article{Jikia:1996bi,
    author = "Jikia, G. and Tkabladze, A.",
    title = "{QCD corrections to heavy quark pair production in polarized gamma gamma collisions and the intermediate mass Higgs signal}",
    eprint = "hep-ph/9601384",
    archivePrefix = "arXiv",
    reportNumber = "JINR-E2-96-31, JINR-PREPRINT-E2-96-31",
    doi = "10.1103/PhysRevD.54.2030",
    journal = "Phys. Rev. D",
    volume = "54",
    pages = "2030--2042",
    year = "1996"
}

@inproceedings{Beneke:2000hk,
    author = "Beneke, M. and others",
    title = "{Top quark physics}",
    eprint = "hep-ph/0003033",
    archivePrefix = "arXiv",
    reportNumber = "CERN-TH-2000-100, FERMILAB-CONF-00-398-T",
    doi = "10.5170/CERN-2000-004.419",
    pages = "419--529",
    month = "3",
    year = "2000"
}

@article{ParticleDataGroup:2024cfk,
    author = "Navas, S. and others",
    collaboration = "Particle Data Group",
    title = "{Review of particle physics}",
    doi = "10.1103/PhysRevD.110.030001",
    journal = "Phys. Rev. D",
    volume = "110",
    number = "3",
    pages = "030001",
    year = "2024"
}

@article{Ginzburg:1981vm,
    author = "Ginzburg, I. F. and Kotkin, G. L. and Serbo, V. G. and Telnov, Valery I.",
    title = "{Colliding gamma e and gamma gamma Beams Based on the Single Pass Accelerators (of Vlepp Type)}",
    reportNumber = "IYF-81-102",
    doi = "10.1016/0167-5087(83)90173-4",
    journal = "Nucl. Instrum. Meth.",
    volume = "205",
    pages = "47--68",
    year = "1983"
}

@article{Ginzburg:1982yr,
    author = "Ginzburg, I. F. and Kotkin, G. L. and Panfil, S. L. and Serbo, V. G. and Telnov, Valery I.",
    title = "{Colliding gamma e and gamma gamma Beams Based on the Single Pass e+ e- Accelerators. 2. Polarization Effects. Monochromatization Improvement}",
    reportNumber = "IYF-82-160",
    doi = "10.1016/0167-5087(84)90128-5",
    journal = "Nucl. Instrum. Meth. A",
    volume = "219",
    pages = "5--24",
    year = "1984"
}

@article{Wootters:1997id,
    author = "Wootters, William K.",
    title = "{Entanglement of formation of an arbitrary state of two qubits}",
    eprint = "quant-ph/9709029",
    archivePrefix = "arXiv",
    doi = "10.1103/PhysRevLett.80.2245",
    journal = "Phys. Rev. Lett.",
    volume = "80",
    pages = "2245--2248",
    year = "1998"
}

@article{Clauser:1969ny,
    author = "Clauser, John F. and Horne, Michael A. and Shimony, Abner and Holt, Richard A.",
    title = "{Proposed experiment to test local hidden variable theories}",
    doi = "10.1103/PhysRevLett.23.880",
    journal = "Phys. Rev. Lett.",
    volume = "23",
    pages = "880--884",
    year = "1969"
}

@article{Horodecki:1995nsk,
    author = "Horodecki, R. and Horodecki, P. and Horodecki, M.",
    title = "{Violating Bell inequality by mixed 
				
					spin-
						
							1
							2
						
					
				 states: necessary and sufficient condition
			}",
    doi = "10.1016/0375-9601(95)00214-N",
    journal = "Phys. Lett. A",
    volume = "200",
    number = "5",
    pages = "340--344",
    year = "1995"
}

@article{Bernreuther:2015yna,
    author = "Bernreuther, Werner and Heisler, Dennis and Si, Zong-Guo",
    title = "{A set of top quark spin correlation and polarization observables for the LHC: Standard Model predictions and new physics contributions}",
    eprint = "1508.05271",
    archivePrefix = "arXiv",
    primaryClass = "hep-ph",
    reportNumber = "TTK-15-16",
    doi = "10.1007/JHEP12(2015)026",
    journal = "JHEP",
    volume = "12",
    pages = "026",
    year = "2015"
}

@article{Bernreuther:2001rq,
    author = "Bernreuther, W. and Brandenburg, A. and Si, Z. G. and Uwer, P.",
    title = "{Top quark spin correlations at hadron colliders: Predictions at next-to-leading order QCD}",
    eprint = "hep-ph/0107086",
    archivePrefix = "arXiv",
    reportNumber = "DESY-01-085, PITHA-01-06, TTP-01-16",
    doi = "10.1103/PhysRevLett.87.242002",
    journal = "Phys. Rev. Lett.",
    volume = "87",
    pages = "242002",
    year = "2001"
}

@article{Horodecki:2009zz,
    author = "Horodecki, Ryszard and Horodecki, Pawel and Horodecki, Michal and Horodecki, Karol",
    title = "{Quantum entanglement}",
    eprint = "quant-ph/0702225",
    archivePrefix = "arXiv",
    doi = "10.1103/RevModPhys.81.865",
    journal = "Rev. Mod. Phys.",
    volume = "81",
    pages = "865--942",
    year = "2009"
}

@article{Gigena:2017agv,
    author = "Gigena, N. and Rossignoli, R.",
    title = "{Bipartite entanglement in fermion systems}",
    eprint = "1704.08091",
    archivePrefix = "arXiv",
    primaryClass = "quant-ph",
    doi = "10.1103/PhysRevA.95.062320",
    journal = "Phys. Rev. A",
    volume = "95",
    pages = "062320",
    year = "2017"
}

@article{Bernreuther:2017yhg,
    author = "Bernreuther, Werner and Galler, Peter and Si, Zong-Guo and Uwer, Peter",
    title = "{Production of heavy Higgs bosons and decay into top quarks at the LHC. II: Top-quark polarization and spin correlation effects}",
    eprint = "1702.06063",
    archivePrefix = "arXiv",
    primaryClass = "hep-ph",
    reportNumber = "HU-EP-17-04, TTK-17-03",
    doi = "10.1103/PhysRevD.95.095012",
    journal = "Phys. Rev. D",
    volume = "95",
    number = "9",
    pages = "095012",
    year = "2017"
}

@article{Maltoni:2024tul,
    author = "Maltoni, Fabio and Severi, Claudio and Tentori, Simone and Vryonidou, Eleni",
    title = "{Quantum detection of new physics in top-quark pair production at the LHC}",
    eprint = "2401.08751",
    archivePrefix = "arXiv",
    primaryClass = "hep-ph",
    doi = "10.1007/JHEP03(2024)099",
    journal = "JHEP",
    volume = "03",
    pages = "099",
    year = "2024"
}

@article{Fabbrichesi:2022ovb,
    author = "Fabbrichesi, Marco and Floreanini, Roberto and Gabrielli, Emidio",
    title = "{Constraining new physics in entangled two-qubit systems: top-quark, tau-lepton and photon pairs}",
    eprint = "2208.11723",
    archivePrefix = "arXiv",
    primaryClass = "hep-ph",
    doi = "10.1140/epjc/s10052-023-11307-2",
    journal = "Eur. Phys. J. C",
    volume = "83",
    number = "2",
    pages = "162",
    year = "2023"
}

@article{Aoude:2025jzc,
    author = "Aoude, Rafael and Banks, Hannah and White, Chris D. and White, Martin J.",
    title = "{Probing new physics in the top sector using quantum information}",
    eprint = "2505.12522",
    archivePrefix = "arXiv",
    primaryClass = "hep-ph",
    doi = "10.1103/c9yb-81kv",
    journal = "Phys. Rev. D",
    volume = "113",
    number = "11",
    pages = "115066",
    year = "2026"
}

@article{Afik:2022dgh,
    author = "Afik, Yoav and de Nova, Juan Ram{\'o}n Mu{\~n}oz",
    title = "{Quantum Discord and Steering in Top Quarks at the LHC}",
    eprint = "2209.03969",
    archivePrefix = "arXiv",
    primaryClass = "quant-ph",
    doi = "10.1103/PhysRevLett.130.221801",
    journal = "Phys. Rev. Lett.",
    volume = "130",
    number = "22",
    pages = "221801",
    year = "2023"
}

@article{Arai:2026jtc,
    author = "Arai, Masato and Mawatari, Kentarou and Okada, Nobuchika",
    title = "{Disentangling new physics with quantum entanglement in $t\bar{t}$ production at future lepton colliders}",
    eprint = "2604.21332",
    archivePrefix = "arXiv",
    primaryClass = "hep-ph",
    month = "4",
    year = "2026"
}

@article{Aoude:2022imd,
    author = "Aoude, Rafael and Madge, Eric and Maltoni, Fabio and Mantani, Luca",
    title = "{Quantum SMEFT tomography: Top quark pair production at the LHC}",
    eprint = "2203.05619",
    archivePrefix = "arXiv",
    primaryClass = "hep-ph",
    reportNumber = "CP3-22-14",
    doi = "10.1103/PhysRevD.106.055007",
    journal = "Phys. Rev. D",
    volume = "106",
    number = "5",
    pages = "055007",
    year = "2022"
}

@article{Subba:2026nzs,
    author = "Subba, Amir and Shi, Yu",
    title = "{Quantumness of top quark pairs produced at LHC within SMEFT framework}",
    eprint = "2605.12033",
    archivePrefix = "arXiv",
    primaryClass = "hep-ph",
    month = "5",
    year = "2026"
}

@article{Czarnecki:2001gi,
    author = "Czarnecki, Andrzej and Melnikov, Kirill",
    title = "{Top quark threshold production at a gamma gamma collider at next-to-next-to-leading order}",
    eprint = "hep-ph/0108233",
    archivePrefix = "arXiv",
    reportNumber = "SLAC-PUB-8972, ALBERTA-THY-13-01",
    doi = "10.1103/PhysRevD.65.051501",
    journal = "Phys. Rev. D",
    volume = "65",
    pages = "051501",
    year = "2002"
}

@article{Capatti:2025khs,
    author = "Capatti, Zeno and Fraaije, Mathijs and Hirschi, Valentin and Huber, Lucien and Ruijl, Ben and Shao, Hua-Sheng",
    title = "{NNLO QCD corrections to {\ensuremath{\gamma}}{\ensuremath{\gamma}} {\textrightarrow} $ Q\overline{Q} $ from Local Unitarity combined with Coulomb resummation and NLO EW effects}",
    eprint = "2509.13439",
    archivePrefix = "arXiv",
    primaryClass = "hep-ph",
    doi = "10.1007/JHEP03(2026)068",
    journal = "JHEP",
    volume = "03",
    pages = "068",
    year = "2026"
}

@article{Denner:1995ar,
    author = "Denner, Ansgar and Dittmaier, S. and Strobel, M.",
    title = "{Radiative corrections to $\gamma \gamma \to t \bar{t}$ in the electroweak standard model}",
    eprint = "hep-ph/9507372",
    archivePrefix = "arXiv",
    reportNumber = "BI-TP-95-27, WUE-ITP-95-017",
    doi = "10.1103/PhysRevD.53.44",
    journal = "Phys. Rev. D",
    volume = "53",
    pages = "44--54",
    year = "1996"
}

@article{Penin:1998mx,
    author = "Penin, A. A. and Pivovarov, A. A.",
    title = "{Analytical results for e+ e- ---{\ensuremath{>}} t anti-t and gamma gamma ---{\ensuremath{>}} t anti-t observables near the threshold up to the next-to-next-to leading order of NRQCD}",
    eprint = "hep-ph/9904278",
    archivePrefix = "arXiv",
    reportNumber = "MZ-TH-98-61",
    doi = "10.1134/1.1349450",
    journal = "Phys. Atom. Nucl.",
    volume = "64",
    pages = "275--293",
    year = "2001"
}

@article{Melles:1998xd,
    author = "Melles, Michael and Stirling, W. James",
    title = "{Heavy quark production at a gamma gamma collider: The Effect of large logarithmic perturbative corrections}",
    eprint = "hep-ph/9810432",
    archivePrefix = "arXiv",
    reportNumber = "DTP-98-66",
    doi = "10.1007/s100529900025",
    journal = "Eur. Phys. J. C",
    volume = "9",
    pages = "101--106",
    year = "1999"
}

@article{Atkinson:2021eox,
    author = "Atkinson, Oliver and Black, Matthew and Lenz, Alexander and Rusov, Aleksey and Wynne, James",
    title = "{Cornering the Two Higgs Doublet Model Type II}",
    eprint = "2107.05650",
    archivePrefix = "arXiv",
    primaryClass = "hep-ph",
    reportNumber = "SI-HEP-2021-20, SFB-257-P3H-21-049",
    doi = "10.1007/JHEP04(2022)172",
    journal = "JHEP",
    volume = "04",
    pages = "172",
    year = "2022"
}

@article{Peskin:1991sw,
    author = "Peskin, Michael E. and Takeuchi, Tatsu",
    title = "{Estimation of oblique electroweak corrections}",
    reportNumber = "SLAC-PUB-5618",
    doi = "10.1103/PhysRevD.46.381",
    journal = "Phys. Rev. D",
    volume = "46",
    pages = "381--409",
    year = "1992"
}

@article{ATLAS:2024vxm,
    author = "Aad, Georges and others",
    collaboration = "ATLAS",
    title = "{Search for heavy neutral Higgs bosons decaying into a top quark pair in 140 fb$^{−1}$ of proton-proton collision data at $ \sqrt{s} $ = 13 TeV with the ATLAS detector}",
    eprint = "2404.18986",
    archivePrefix = "arXiv",
    primaryClass = "hep-ex",
    reportNumber = "CERN-EP-2024-090",
    doi = "10.1007/JHEP08(2024)013",
    journal = "JHEP",
    volume = "08",
    pages = "013",
    year = "2024"
}

@article{Brandenburg:2005uu,
    author = "Brandenburg, A. and Si, Z. G.",
    title = "{Top quark pair production and decay at a polarized photon collider}",
    eprint = "hep-ph/0503153",
    archivePrefix = "arXiv",
    reportNumber = "DESY-05-043, SDU-HEP200501",
    doi = "10.1016/j.physletb.2005.04.003",
    journal = "Phys. Lett. B",
    volume = "615",
    pages = "68--78",
    year = "2005"
}

@article{Branco:2011iw,
    author = "Branco, G. C. and Ferreira, P. M. and Lavoura, L. and Rebelo, M. N. and Sher, Marc and Silva, Joao P.",
    title = "{Theory and phenomenology of two-Higgs-doublet models}",
    eprint = "1106.0034",
    archivePrefix = "arXiv",
    primaryClass = "hep-ph",
    doi = "10.1016/j.physrep.2012.02.002",
    journal = "Phys. Rept.",
    volume = "516",
    pages = "1--102",
    year = "2012"
}

@article{Eriksson:2009ws,
    author = "Eriksson, David and Rathsman, Johan and Stal, Oscar",
    title = "{2HDMC: Two-Higgs-Doublet Model Calculator Physics and Manual}",
    eprint = "0902.0851",
    archivePrefix = "arXiv",
    primaryClass = "hep-ph",
    doi = "10.1016/j.cpc.2009.09.011",
    journal = "Comput. Phys. Commun.",
    volume = "181",
    pages = "189--205",
    year = "2010"
}

@article{Bernon:2015wef,
    author = "Bernon, J{\'e}r{\'e}my and Gunion, John F. and Haber, Howard E. and Jiang, Yun and Kraml, Sabine",
    title = "{Scrutinizing the alignment limit in two-Higgs-doublet models. II. m$_H$=125  GeV}",
    eprint = "1511.03682",
    archivePrefix = "arXiv",
    primaryClass = "hep-ph",
    doi = "10.1103/PhysRevD.93.035027",
    journal = "Phys. Rev. D",
    volume = "93",
    number = "3",
    pages = "035027",
    year = "2016"
}

@article{Erdogan:2026cbe,
    author = "Erdogan, Eren and Mohan, Kirtimaan A. and Xie, Keping and Yanez, Marcel",
    title = "{Quantum Tomography of Top Quarks as a Probe of Charge-Parity Violation}",
    eprint = "2607.28761",
    archivePrefix = "arXiv",
    primaryClass = "hep-ph",
    reportNumber = "MSUHEP-26-011",
    month = "7",
    year = "2026"
}

@article{CMS:2025dzq,
    author = "Hayrapetyan, Aram and others",
    collaboration = "CMS",
    title = "{Search for heavy pseudoscalar and scalar bosons decaying to a top quark pair in proton{\textendash}proton collisions at $\sqrt{s} = 13\,\textrm{TeV}$}",
    eprint = "2507.05119",
    archivePrefix = "arXiv",
    primaryClass = "hep-ex",
    reportNumber = "CMS-HIG-22-013, CERN-EP-2025-124",
    doi = "10.1088/1361-6633/ae2207",
    journal = "Rept. Prog. Phys.",
    volume = "88",
    number = "12",
    pages = "127801",
    year = "2025"
}

@article{Xu:2025vmy,
    author = "Xu, Haotian and Wang, Yufei and Han, Xiao-Fang and Wang, Lei",
    title = "{95 GeV Higgs boson and nano-Hertz gravitational waves from domain walls in the next-to-two-Higgs-doublet model}",
    eprint = "2505.03592",
    archivePrefix = "arXiv",
    primaryClass = "hep-ph",
    doi = "10.1088/1674-1137/ae2082",
    journal = "Chin. Phys. C",
    volume = "50",
    number = "1",
    pages = "013108",
    year = "2026"
}

@article{Aoki:2009ha,
    author = "Aoki, Mayumi and Kanemura, Shinya and Tsumura, Koji and Yagyu, Kei",
    title = "{Models of Yukawa interaction in the two Higgs doublet model, and their collider phenomenology}",
    eprint = "0902.4665",
    archivePrefix = "arXiv",
    primaryClass = "hep-ph",
    reportNumber = "TU-839, UT-HET-022, IC-2009-007",
    doi = "10.1103/PhysRevD.80.015017",
    journal = "Phys. Rev. D",
    volume = "80",
    pages = "015017",
    year = "2009"
}

@article{Jiang:2026sfw,
    author = "Jiang, Jun and Si, Zong-Guo and Zhang, Han and Zhang, Xin-Yi",
    title = "{Spin correlations and quantum entanglement in $γγ\to t\bar t$ at polarized photon colliders with NLO QCD corrections}",
    eprint = "2606.21152",
    archivePrefix = "arXiv",
    primaryClass = "hep-ph",
    month = "6",
    year = "2026"
}

@article{Kling:2018xud,
    author = "Kling, Felix and Li, Honglei and Pyarelal, Adarsh and Song, Huayang and Su, Shufang",
    title = "{Exotic Higgs Decays in Type-II 2HDMs at the LHC and Future 100 TeV Hadron Colliders}",
    eprint = "1812.01633",
    archivePrefix = "arXiv",
    primaryClass = "hep-ph",
    reportNumber = "UCI-TR-2018-09",
    doi = "10.1007/JHEP06(2019)031",
    journal = "JHEP",
    volume = "06",
    pages = "031",
    year = "2019"
}

\end{document}